# Comparing Femtosecond Optical Tweezers with Conventional CW Optical Tweezers

AJITESH SINGH,[1] KRISHNA KANT SINGH,[1] DEEPAK KUMAR,[1] DEBABRATA GOSWAMI[1, 2]

[1]Department of Chemistry, Indian Institute of Technology Kanpur, Kanpur 208016, India,
[2]Centre of Lasers and Photonics, Indian Institute of Technology Kanpur, Kanpur 208016, India

**In this work, we present a comparative study between continuous-wave (CW) and pulsed optical tweezers for 250 nm, 500 nm and 1 µm radius polystyrene beads at 5 different laser powers. We have used a Ti:Sapphire (MIRA 900F) laser that can be easily switched from CW to pulsed mode of operation, so there is no change in the experimental conditions in the two cases.  We have measured the difference in the trap strength in both the cases by fitting the power spectrum curve with Lorentzian. As it turns out, trapping with pulsed tweezers seems to be more effective for the smaller particles and as the particle size is increased both CW and pulsed tweezers appear to be equally effective at lower average laser powers but as the power is increased pulsed tweezers do a better job at stable trapping.**

From early tweezers days to approximately the next 25 years tweezing was primarily done with CW lasers. And although there are many key elements at the heart of a tweezer's setup, choice of laser wavelength is extremely important for minimizing the optical damage; especially while working with biological samples (one does certainly try not to fry the sample). So, usually the most common light source for trapping had been CW near IR lasers. However, since the early 2000s, pulsed and broadband lasers have been used more frequently in place of or in addition to CW lasers. This would make it possible to use spectroscopy along with optical trapping to detect nonlinear effects. Pulsed lasers enable access to multiphoton processes that would otherwise demand very high average powers of CW. When using ultrashort pulsed lasers, which typically have pulse durations of 100 femtoseconds or fewer, these effects become dramatically stronger. The idea that trapping could be done using pulsed lasers was first experimentally showed by Malmqvist and Hertz [1] and it was Dholakia et al. who were the first one to use femtosecond lasers for trapping and manipulating micron sized particles with it [2-4]. It has been suggested that, if the nonlinear effects were to be neglected, in Rayleigh [5, 6] and Ray optics regime [7], similar forces will be exerted on the trapped particle from CW and pulse beam of same average laser power. Although, in Lorentz-Mie regime, where the forces depend on the ratio of particle size to that of the wavelength [8-10], the nature of the forces might be different even without considering the nonlinear effects. Effect of the pulse duration on optical trapping with an ultrashort pulsed laser with high repetition rate were examined by Shane et al. [11] using numerical simulations based on GLMT and experiments, and found that there was no significant effect of the laser pulse duration on optical trapping. However, work by Wang et al. [12] and Preez-Wilkinson et al. [13] shows that there will be an increase in the radiation force, in both longitudinal and transverse components, as the laser pulse is decreased. A different experimental approach was adopted by Liu et al. [14] for comparison of CW and fs tweezers, based on the dynamics of trapped polystyrene sphere and suggested that particle had higher trapping stability under CW beam than fs pulse train. In this letter, we have used power spectrum method [15] for measuring the trap stiffness associated, when trapping was done in CW and pulsed mode. Our Ti:Sapphire (MIRA 900F) laser is just the perfect light source, as it can be switched between pulsed and CW mode of operation just by a flick of a switch. We have performed and presented results for trapping of 250 nm, 500 nm and 1µm radius polystyrene beads under both the conditions. The corner frequencies are intrinsically related to the trap stiffness, so by measuring the variation in the corner frequencies one could get a feel of how corresponding trap stiffness would change. We have trapped 250 nm beads from 25 mW to 45 mW, and 500 nm and 1 µm beads from 07 mW to 27 mW. Based on our results, we have found that for smaller size particle (250 nm) the pulsed lasers seem to have a clear upper hand in terms of trap stiffness, whereas for intermediate size particle (500 nm) at lower powers both CW and pulsed tweezers appear to have similar strength, however, at higher powers pulsed tweezers would have upper hand. For large size particle (1µm) it was hard to comment on a clear winner between the two modes, however, the slope suggests that pulsed tweezers will again do a better job at higher average laser powers.

Figure 1 is the schematic of optical tweezers setup. With a combination of a half-wave plate and a polarizing beam-splitter we precisely control the laser power reaching the sample chamber. The laser beam's initial diameter is not large enough to overfill the back aperture of the objective lens, so a beam expander is placed in the light path. The beam is then reflected off a dichroic mirror to the objective lens which focuses the light beam to trap polystyrene beads. The beads are transparent, and the transmitted light is then collected and collimated by another objective lens (condenser) which is then focused and incident on a quadrant photodiode (QPD). The QPD is connected with an oscilloscope which in turn is connected to a personal computer (PC) via GPIB card. We acquire the QPD data using a LabVIEW program. The polystyrene beads used here are dye coated and when trapping is done in pulsed condition, they generate two-photon fluorescence (TPF). The TPF generated is collected and collimated by the focusing objective and can be seen on the CCD camera, which is also connected to the PC. The zoomed-in part shows a cartoon picture of the sample chamber.

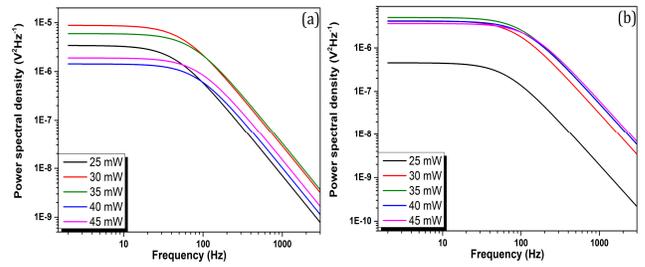

Fig. 2. Lorentzian fit curves for 250 nm radius particle at all the powers. (a) For CW case. (b) For ML case.

It is evident from values documented in table 1 that, as the laser power is increased there is a linear increase in the corner frequency values for both CW and pulsed. Whereas, if we compare the corner frequencies for CW case at each power with that for the pulsed case, we find that the pulsed corner frequencies are always higher. As trap stiffness is directly proportional to the corner frequency, trap stiffness for pulsed case is always higher than CW case at each average laser power.

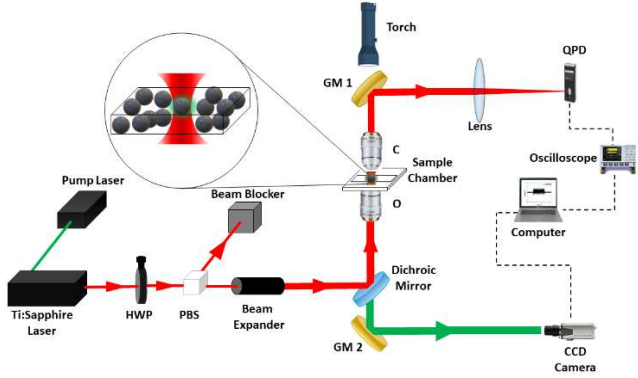

Fig. 1. Optical tweezers setup. HWP: Half-wave plate, PBS: Polarizing Beam-Splitter, GM: Gold Mirror, O: Objective lens, C: Condenser lens, QPD: Quadrant Photodiode.

Brownian motion of a spherical particle trapped in a harmonic potential is given by the Langevin equation as [15]

$$m\ddot{x}(t) + \gamma_0 \dot{x}(t) + \kappa x(t) = (2k_B T \gamma_0)^{1/2} \quad (1)$$

Where, $m$ is mass of the particle, $\gamma_0$ is the friction coefficient, $\kappa$ is the trap stiffness, $k_B$ is the Boltzmann constant, $T$ is the absolute temperature, $x(t)$ is the particle's position as a function of time. The term on the right-hand side represents a random process with Gaussian distribution. Raw data from the quadrant photodiode (QPD) is analyzed using power spectrum method and the corresponding power spectrums are fitted with the Lorentzian [15] to obtain the corner frequency. Corner frequency is directly related to the trap stiffness as [15]

$$\kappa = 12\pi^2 f_c \eta r \quad (2)$$

Where $r$ is the radius of the trapped particle, $f_c$ is the corner frequency and $\eta$ is viscosity of the surrounding medium. So, higher the corner frequency associated with the trapped particle, higher will be the trap stiffness.

Figure 2 (a) and (b), represents the consolidated Lorentzian fits for CW and pulsed mode respectively (ML stands for mode-locked). Specific power spectrum curves at each power can be found in the supplimentary information. The corner frequency values associated with each curve is documented in table 1 and with respect to each corner frequency corresponding trap stiffness in calculated using equation (2). Viscosity of water is taken from the literature.

| Power (mW) | CW | | ML | |
| --- | --- | --- | --- | --- |
| | Corner frequency (Hz) | Trap Stiffness (fN/nm) | Corner frequency (Hz) | Trap Stiffness (fN/nm) |
| 25 | 48.62 | 1.28 | 65.50 | 1.73 |
| 30 | 59.17 | 1.56 | 83.72 | 2.21 |
| 35 | 74.02 | 1.95 | 103.31 | 2.72 |
| 40 | 84.53 | 2.22 | 111.53 | 2.94 |
| 45 | 89.39 | 2.36 | 127.53 | 3.36 |

Table 1. Consolidated data for 250 nm radius beads.

Figure 3 represents variation of the corener frequency with laser power. Slope for the pulsed case is 2.80, whereas for CW case its 2.05. So, as the laser power is increased particles trapped with pulsed tweezers will be more and more stable.

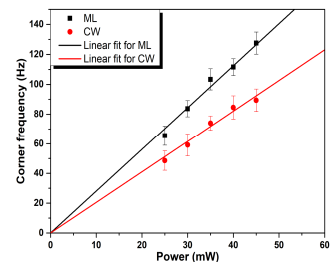

Fig. 3. Linear plot showing variation in corner frequencies with respect to average laser power.

Figure 4 (a) and (b) represents the Lorentzian fits for the 500 nm radius particle for CW and pulsed case respectively. And the corner frequency values associated with each curve are listed in table 2. Similar to the earlier case, as the laser power is increased there is a corresponding increase in the corner frequency value for both pulsed and CW case, however, the corner frequencies for pulsed case are again higher than their corresponding CW counterpart. Trap stiffness, as a consquence, are also higher for the pulsed case at each power.

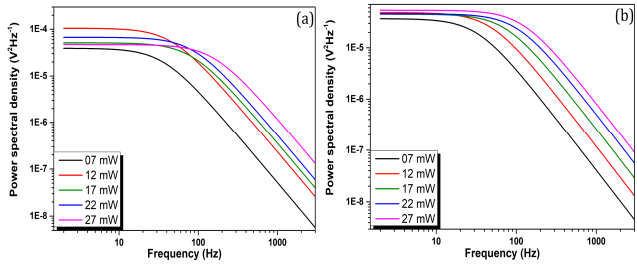

Fig. 4. Lorentzian fit curves for 500 nm radius particle at all the powers. (a) For CW case. (b) For ML case.

| Power (mW) | CW | | ML | |
|---|---|---|---|---|
| | Corner frequency (Hz) | Trap Stiffness (fN/nm) | Corner frequency (Hz) | Trap Stiffness (fN/nm) |
| 07 | 33.44 | 0.88 | 38.94 | 1.02 |
| 12 | 51.34 | 1.35 | 66.90 | 1.76 |
| 17 | 68.18 | 1.80 | 76.05 | 2.00 |
| 22 | 89.18 | 2.35 | 105.87 | 2.79 |
| 27 | 112.62 | 2.97 | 137.07 | 3.53 |

Table 2. Consolidated data for 500 nm radius beads.

Figure 5 represents the variation in the corner frequencies for 500 nm radius beads with respect to average laser power. Slope for the pulsed case is 4.87 and that for CW case in 4.11, so again as in previous case the pulsed tweezers will do a better job of stable trapping than CW tweezers at higher laser powers. However at lower powers both techniques do comparble job.

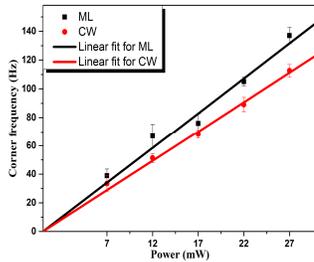

Fig. 5. Linear plot showing variation in corner frequencies with average laser power.

Figure 6 (a) and (b) represents the Lorentzian fits for 1μm radius beads for CW and pulsed case repectively. The corresponding corner frequency values are recorded in table 3, as in the earlier cases the corner frequency values individually for CW and pulsed mode increased as the average laser power is increased.

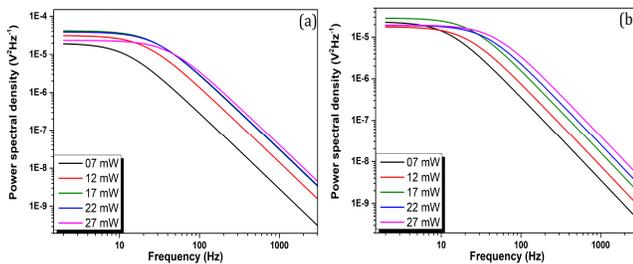

Fig. 6. Lorentzian fit curves for 1um radius particle at all the powers. (a) For CW case. (b) For ML case.

Unlike earlier cases, for 1 μm radius particle the corner frequencies at 12 mW and 17 mW are higher in CW case than in pulsed case and so is the corresponding trap stiffness.

| Power (mW) | CW | | ML | |
|---|---|---|---|---|
| | Corner frequency (Hz) | Trap Stiffness (fN/nm) | Corner frequency (Hz) | Trap Stiffness (fN/nm) |
| 07 | 11.98 | 0.32 | 13.43 | 0.33 |
| 12 | 20.20 | 0.53 | 21.95 | 0.55 |
| 17 | 25.19 | 0.66 | 24.06 | 0.63 |
| 22 | 30.03 | 0.79 | 36.67 | 0.96 |
| 27 | 38.43 | 1.10 | 45.88 | 1.21 |

Table 3. Consolidated data for 1 μm radius beads.

Figure 7 represents the variation of corner frequencies for 1 μm radius beads with respect to the average laser powers. In this case, it is hard to comment on whether CW is better or pulsed because throughout the power range the error bars for both the cases overlap (barring at 22 mW). Although the slope for pulsed case (1.63) is still higher than for CW (1.44) suggesting that at higher powers pulsed would still do a better job at trapping the particle than CW.

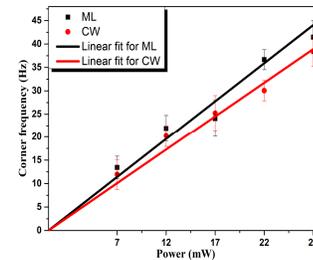

Fig. 7. Linear plot showing variation in the corner frequencies with respect to the average laser power.

Interestingly, the results that we have got for three particles of different sizes can be very well understood with the help of a theoretical work by Bandyopadhyay et al. [16]. What they have theoretically predicted is that when the laser is operating in CW mode there always are some inherent, unavoidable, heating effects associated with it, which at focus will relatively change the refractive index of the medium in which the beads are suspended (water in this case). This change in the refractive index results in an increase in the focus spot and this increase then in turn decreases the intensity gradient at the focus, which eventually decrease the gradient force. Therefore, CW tweezers could have trapped particles relatively more strongly if not for these inherent heating effects associated with CW laser beam. Whereas, when the laser is operating in pulsed mode there is delay between the two successive pulses. In our case, since our Ti:Sapphire laser has a repetition rate of 76 MHz, which means the interacting system has ~13 ns to relax before the next pulse hits it and therefore the cumulative heating with pulsed laser is always less than that due to the CW laser. On top of having relatively lesser heating effects, pulsed laser will also have extremely high peak powers associated with them, which are capable of triggering Kerr effect in the medium they interact with (polystyrene beads and water in our case). Whereas the peak powers in CW case are not high enough to generate any Kerr effect. The nonlinear refractive index of polystyrene bead is 5.17 ×10$^{-17}$ m$^2$/W and it was argued in [16] that the Kerr effect increases the total force by a factor of 10, although some heating effects will tend reduce the overall impact. Now, this change in the focal spot is relatively large for a 250 nm bead or in other terms 250 nm bead will be more sensitive to this change in the focal spot than the 500 nm and 1 µm beads. And probably this is why the difference in pulsed and CW trapping were more pronounced for 250 nm radius beads and as the bead size was increased to 500 nm radius, pulsed and CW trapping appeared to be similar at the lower average laser powers, whereas when the particle size was changed to 1 µm, this change in the focal spot was almost negligible and therefore for 1 µm radius beads the trapping with CW and pulsed appeared to do similar job.

## Acknowledgments

We acknowledge the funding support of this research from MEITY, SERB, and STC-ISRO of the Govt. of India. A. Singh and D. Kumar thanks UGC India, and K. K. Singh thanks IIT Kanpur for their respective graduate fellowships. All authors thank S. Goswami for language correction and editing.

## References

[1] L. Malmqvist and H. M. Hertz, "Second-harmonic generation in optically trapped nonlinear particles with pulsed lasers," *Appl. Opt.,* vol. 34, no. 18, pp. 3392-3397, 1995/06/20 1995, doi: 10.1364/AO.34.003392.

[2] B. Agate, C. Brown, W. Sibbett, and K. Dholakia, "Femtosecond optical tweezers for in-situ control of two-photon fluorescence," *Opt. Express,* vol. 12, no. 13, pp. 3011-3017, 2004.

[3] K. Dholakia *et al.*, "Imaging in optical micromanipulation using two-photon excitation," *New J. Phys.,* vol. 6, no. 1, p. 136, 2004.

[4] H. Little, C. Brown, V. Garcés-Chávez, W. Sibbett, and K. Dholakia, "Optical guiding of microscopic particles in femtosecond and continuous wave Bessel light beams," *Opt. Express,* vol. 12, no. 11, pp. 2560-2565, 2004.

[5] J.-l. Deng, Q. Wei, Y.-z. Wang, and Y.-q. Li, "Numerical modeling of optical levitation and trapping of the "stuck" particles with a pulsed optical tweezers," *Opt. Express,* vol. 13, no. 10, pp. 3673-3680, 2005.

[6] A. Usman, W.-Y. Chiang, and H. Masuhara, "Femtosecond trapping efficiency enhanced for nano-sized silica spheres," in *Optical Trapping and Optical Micromanipulation IX*, 2012, vol. 8458: SPIE, pp. 504-510.

[7] Q. Xing, F. Mao, L. Chai, and Q. Wang, "Numerical modeling and theoretical analysis of femtosecond laser tweezers," *Opt. Laser Technol.,* vol. 36, no. 8, pp. 635-639, 2004.

[8] P. M. Neto and H. Nussenzveig, "Theory of optical tweezers," *Europhys. Lett.,* vol. 50, no. 5, p. 702, 2000.

[9] A. Mazolli, P. Maia Neto, and H. Nussenzveig, "Theory of trapping forces in optical tweezers," *Proc. R. Soc. Lond. A,* vol. 459, no. 2040, pp. 3021-3041, 2003.

[10] A. B. Stilgoe, T. A. Nieminen, G. Knöner, N. R. Heckenberg, and H. Rubinsztein-Dunlop, "The effect of Mie resonances on trapping in optical tweezers," *Opt. Express* vol. 16, no. 19, pp. 15039-15051, 2008.

[11] J. C. Shane, M. Mazilu, W. M. Lee, and K. Dholakia, "Effect of pulse temporal shape on optical trapping and impulse transfer using ultrashort pulsed lasers," *Opt. Express,* vol. 18, no. 7, pp. 7554-7568, 2010.

[12] L.-G. Wang and C.-L. Zhao, "Dynamic radiation force of a pulsed Gaussian beam acting on a Rayleigh dielectric sphere," *Opt. Express,* vol. 15, no. 17, pp. 10615-10621, 2007.

[13] N. du Preez-Wilkinson, A. B. Stilgoe, T. Alzaidi, H. Rubinsztein-Dunlop, and T. A. Nieminen, "Forces due to pulsed beams in optical tweezers: linear effects," *Opt. Express,* vol. 23, no. 6, pp. 7190-7208, 2015.

[14] T.-H. Liu, W.-Y. Chiang, A. Usman, and H. Masuhara, "Optical trapping dynamics of a single polystyrene sphere: continuous wave versus femtosecond lasers," *J. Phys. Chem. C,* vol. 120, no. 4, pp. 2392-2399, 2016.

[15] K. Berg-Sørensen and H. Flyvbjerg, "Power spectrum analysis for optical tweezers," *Rev. Sci. Instrum.,* vol. 75, no. 3, pp. 594-612, 2004.

[16] S. N. Bandyopadhyay, T. Gaur, and D. Goswami, "Comparative study of the real-time optical trapping in the Rayleigh regime for continuous and femtosecond pulsed lasers," *Optics & Laser Technology,* vol. 136, p. 106770, 2021.